# 100 GHz Transistors from Wafer Scale Epitaxial Graphene

Y.-M. Lin*, C. Dimitrakopoulos, K. A. Jenkins, D. B. Farmer,
H.-Y. Chiu, A. Grill, and Ph. Avouris*

IBM T. J. Watson Research Center, Yorktown Heights, NY 10598

**Abstract:** High-performance graphene field-effect transistors have been fabricated on epitaxial graphene synthesized on a two-inch SiC wafer, achieving a cutoff frequency of 100 GHz for a gate length of 240 nm. The high-frequency performance of these epitaxial graphene transistors not only shows the highest speed for any graphene devices up to date, but it also exceeds that of Si MOSFETs at the same gate length. The result confirms the high potential of graphene for advanced electronics applications, marking an important milestone for carbon electronics.

As the thinnest possible electronic material, merely one atom thick, with very high carrier mobility, graphene offers great potential to create the smallest and fastest transistors among all semiconductor materials. Proof-of-concept demonstration of graphene-based electronics has been provided by demonstrating DC operation of field-effect transistors (FETs) – the fundamental building block of modern microelectronics – using graphene flakes extracted from natural graphite (1), graphene films produced by decomposition of the surface of silicon carbide (SiC) substrates (2) or by chemical vapor deposition of hydrocarbons on catalytic metal surfaces (3). In spite of the high hopes and claims for the debut of the era of carbon electronics over the last decade, the missing critical tests for evaluating the viability of this new material for practical applications lie in the challenges of demonstrating high-frequency graphene transistors, and their compatibility with wafer-scale fabrication that would enable circuit integration.

This work presents such high-performance RF (radio frequency) field-effect transistors fabricated on a two-inch graphene wafer (Fig. 1A) with which a cutoff frequency as high as 100 GHz (FIG. 1D) is achieved. As this was achieved using wafer-scale processing, it represents an important milestone towards the ultimate goal of terahertz integrated circuits using graphene. The results not only demonstrate the highest speed for graphene transistors up to date, but also verify the outstanding electronic properties in graphene materials regardless of their origin.

In this work, graphene was epitaxially grown on the Si face of a high-purity semi-insulating SiC 4H(0001) wafer by thermal decomposition at 1450ºC, yielding a film of 1-2 layers of graphene over the entire wafer. The as-grown graphene film possesses an electron ($n$-type) carrier density of ~ $3\times10^{12}$ cm$^{-2}$ and a Hall-effect mobility between 1000–1500 cm$^2$/V·s. While the carrier mobility in graphene can be, in principle, very high due to the suppression of back-scattering based on pseudo-spin conservation for chiral carriers, the mobility in actual graphene devices is highly sensitive to the environment and may suffer significant degradation by the deposition of the gate insulator. In order to preserve the intrinsic mobility of graphene in the top-gated device structure, an interfacial polymer layer (about 9 ± 3 nm thick) was spin-coated on the graphene prior to the oxide deposition (4), on which a high-$\kappa$ film of $HfO_2$ (10 nm thick) was formed by atomic layer deposition (ALD). The carrier mobility of top-gated Hall bar devices varied between 900-1520 $cm^2/V \cdot s$ across the 2" wafer, indicating that little degradation in graphene mobility, if any, was introduced during the fabrication processes.

In order to evaluate the high frequency response of graphene films grown by epitaxy, arrays of top-gated FETs were fabricated with gate lengths as short as 240 nm. FIG. 1A shows a scanning electron microscope image and a cross-sectional schematic view of such a device. FIG. 1B shows the drain current of a graphene FET measured as a function of gate voltage $V_G$, exhibiting characteristics of an *n*-type transistor where the transport is dominated by electrons. For all graphene FETs of this study, the Dirac point, characterized by a current minimum as a function of gate voltage $V_G$, always occurs at a gate voltage less than -3.5 V. This corresponds to a rather high electron density of greater than $4.0 \times 10^{12}$ $cm^{-2}$ in the graphene channel at a zero gate bias state. Such high electron doping concentration is typical of graphene fabricated from the Si face of a SiC wafer, and is advantageous for achieving low series resistance of graphene FETs by enhancing the conductivity in the un-gated graphene regions between the source and drain contacts. As a result, the device transconductance $g_m$, defined by $dI_D/dV_G$, is nearly constant over a wide $V_G$ range in the ON state (right axis in FIG. 1B). The graphene FET exhibits output characteristics (FIG. 1C) different from those of conventional Si-MOSFETs due of the absence of a bandgap in graphene. While the drain current in the graphene FET possesses a sub-linear $I_D$-$V_D$ dependence as the drain bias increases, no clear current saturation is observed at drain bias up to 2 V or before the device breakdown. This output characteristic leads to a device transconductance that increases with drain voltage for graphene FETs.

The scattering (S) parameters of these transistors were measured to investigate their high-frequency performance. In this respect, the insulating SiC wafer provides an ideal platform for building and characterizing high-frequency devices and circuits by reducing the parasitic capacitance between contacts and interconnects. FIG. 1B shows the short-circuit current gain $|h_{21}|$, which is as the ratio of small-signal drain and gate currents and is derived from the measured S-parameters, as a function of frequency for graphene FETs of two gate lengths, 240 nm and 550nm. The measured current gain displays the $1/f$ frequency dependence expected for an ideal FET, from which a well-defined cutoff frequency, $f_T$ can be obtained as the frequency at which the current gain becomes unity. The magnitude of $f_T$ signifies the highest frequency at which a transistor can propagate an electrical signal, and is the most fundamental figure of merit to benchmark transistor performance. For a gate length of 550 nm, the measured $f_T$ ranges between 20 to 53 GHz. For a shorter gate length of 240 nm, $f_T$ as high as 100 GHz was measured at a drain bias of 2.5 V. This 100 GHz cutoff frequency is the highest speed achieved to date for any type of graphene devices, including exfoliated and CVD-grown graphene. It is also remarkable that the graphene FETs demonstrated here possess a cutoff frequency exceeding that of state-of-the-art Si MOSFETs of the same gate length (~ 40 GHz at 240 nm) (5). In addition to the current gain, the graphene FETs also possess power gain $G_{MAG}$ up to $f_{MAX}$ ~ 14 GHz and 10 GHz for 550-nm and 240-nm gate lengths (6), respectively. Both $f_T$ and $f_{MAX}$ are important figures of merit of transistor performance. $f_T$ reflects intrinsic behavior of a transistor channel, whereas $f_{MAX}$ also strongly depends on other factors such as the device layout, and can be further enhanced, for example, by optimizing the gate contact leads.

The high-performance graphene RF transistors, demonstrated herein using wafer-scale graphene synthesis and conventional device fabrication processes, clearly demonstrate the high potential of graphene for advanced electronics applications.

**FIGURE CAPTION**

(A) Scanning electron microscope image and the schematic cross-sectional view of a top-gated graphene field-effect transistor. The optical image of the two-inch graphene/SiC wafer with arrays of graphene devices is shown on the right. The transistors possess dual-gate channels to increase the drive current and lower the gate resistance. The scale bar is 2 μm. (B) The drain current $I_D$ of a 240-nm-gate-length graphene transistor as a function of gate voltage $V_G$ at drain bias of 1V with the source electrode grounded. The current shown was normalized with respect to the total channel width. The device conductance $g_m = dI_D/dV_G$ is shown on the right axis. (C) The measured drain current $I_D$ as a function of drain bias of a graphene FET with a gate length of 240 nm for various top-gate voltages. (D) Measured small-signal current gain $|h_{21}|$ as a function of frequency $f$ for a 240-nm-gate and a 550-nm-gate graphene FET, represented by (◊) and (∆), respectively. The current gain for both devices exhibits the $1/f$ dependence, where a well-defined cutoff frequency $f_T$ can be determined to be 53 GHz and 100 GHz for the 550-nm and 240-nm devices, respectively.

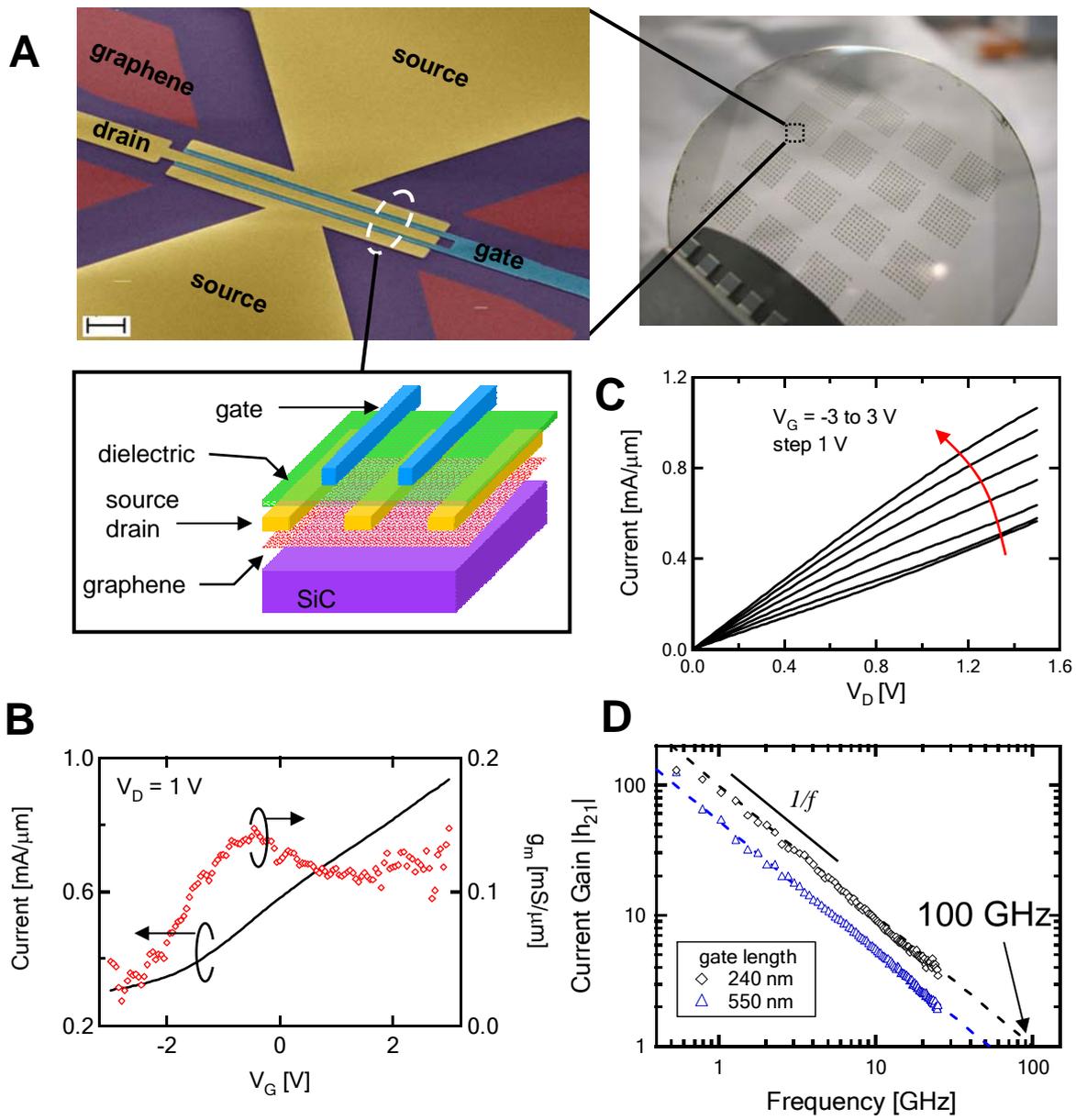

**FIG. 1**